\documentclass[final,5p,times,twocolumn]{elsarticle}

\usepackage{graphicx}
\usepackage{amssymb}
\usepackage{amsmath}
\usepackage{bm}

\usepackage[table]{xcolor}
\usepackage{array,ragged2e}

\usepackage{caption}

\journal{Acta Astronautica}

\begin{document}

\begin{frontmatter}


  \title{Joint analysis of two-way laser range and one-way frequency
    measurements for the gravitational redshift experiment with the
    RadioAstron spacecraft}



\author[kiam,gaish]{M.~V.~Zakhvatkin\corref{mycorrespondingauthor}}
\cortext[mycorrespondingauthor]{Corresponding author at: Keldysh Institute for Applied Mathematics, Russian Academy of Sciences, Miusskaya sq. 4, 125047 Moscow, Russia.}
\ead{zakhvatkin@kiam1.rssi.ru}
\author[gaish,asc]{D.~A.~Litvinov}

\address[kiam]{Keldysh Institute for Applied Mathematics, Russian Academy of Sciences, Miusskaya sq. 4, 125047 Moscow, Russia}

\address[asc]{Astro Space Center of Lebedev Physical Institute of RAS, 84/32
  Profsoyuznaya st., Moscow, Russia 117997}

\address[gaish]{Sternberg Astronomical Institute, Lomonosov Moscow State
  University, Universitetsky pr. 13, 119991 Moscow, Russia}

\begin{abstract}
We consider the problem of joint analysis of two-way laser range and one-way frequency measurements in high-precision tests of general relativity with spacecrafts. Of main interest to such tests is the accuracy of the computed values of the one-way frequency observables. We identify the principal sources of error in these observables to be the errors in the modeled corrections
due to various ``small'' effects, such as that of the troposphere,
the error in the reflection time of the laser pulse from the spacecraft, and the error of fitting the spacecraft trajectory to the laser data. We suggest ways to evaluating these errors.
\end{abstract}


\end{frontmatter}


\section{Introduction}

In this paper we consider the problem of using two-way laser range measurements to increase the accuracy of high-precision space experiments, which use one-way
frequency measurements
of the spacecraft downlink signal as their primary science data. A typical example
of experiments of this kind are tests of the gravitational time dilation, such as Gravity Probe A \cite{vessot-levine-1980-prl} or the gravitational
redshift experiment with the RadioAstron space radio telescope \cite{litvinov-2018-pla}. In such experiments, the accuracy of computed values of the one-way frequency observables  is usually limited by the uncertainty in the spacecraft trajectory and some means needs to be employed to overcome this limitation \cite{vessot-levine-1979-grg}. We suggest using two-way laser range measurements for this purpose. 

The major obstacle to a practical realization of this idea is that the two-way laser range and one-way frequency observables are of substantially
different nature. This makes it difficult to obtain the gravitational frequency
shift as a simple function of the frequency of the received signal, the laser pulse round-trip time and other kinematic parameters, such as the tropospheric delay and the possible offset of the on-board frequency standard's output
frequency from its nominal value. The goal of our research is to identify possible sources of error inherent to the joint analysis of such data and make an attempt to find ways to estimate the magnitude of these errors.

\section{Computed values of frequency observables}
In order to approach the problem of the joint analysis of two-way laser range and one-way frequency measurements, which are two substantially different types of observables, we first consider the models for their computed values.

Let us start with the frequency observable since it is basic to the experiments
we are interested in. By the frequency observable we understand the phase of the signal received
by a station and accumulated during some specified time interval. Below we will be using the following notation for the time tags. Subscripts are used to indicate any of the following three events: 1 -- the event of signal transmission by the station; 2 -- the event of signal transmission, reflection or retransmission by the spacecraft; 3 -- the event of signal reception by the station. The superscript, if present, indicates either the beginning ($b$) or  ending ($e$) of a time interval. The argument of a time variable specifies its time scale. If the argument is not provided, we will assume the time is given in the reference frame used to describe the motion of
the spacecraft, and we will call this the ephemeris time scale. Position vectors, $\mathbf{r}_i$, with the possible values of $i$ being 1, 2, and 3,
will specify, respectively: the position of the ground station reference point at the time of transmission; the position of the spacecraft center of mass at the time of retransmission (for the two-way mode) or transmission (for the one-way mode); the position of the reference point of the station at the time of  reception of the signal:
\begin{align*}
  \mathbf{r}_1 &= \mathbf{r}_{ST}(t_1),\\
  \mathbf{r}_2 &= \mathbf{r}_{KA}(t_2),\\
  \mathbf{r}_3 &= \mathbf{r}_{ST}(t_3).
\end{align*}

Using this notation we can specify  the accumulation time in the time scale
of the station as $(t_3^b(ST), t_3^e(ST))$. The computed value of this observable is:
\begin{align}
  \label{eq:f_c}
  f_c = \frac{1}{T_c}\int_{t_3^b(ST)}^{t_3^e(ST)} f_R dt_3(ST),
\end{align}
where $t(ST)$ is the time given in the ground station time scale, $T_c = t_3^e(ST) - t_3^b(ST)$ is the duration of the accumulation time measured at the station, $f_R$ is the frequency of the received signal. The integral in Eq.~\eqref{eq:f_c} gives the change of the received signal phase over the accumulation time. In the course of the transmission of this signal by the
spacecraft its phase also changed by this same amount, which occurred during $(t_2^b(KA), t_2^e(KA))$:
\begin{align*}
  \int_{t_3^b(ST)}^{t_3^e(ST)} f_R dt_3(ST) =\int_{t_2^b(KA)}^{t_2^e(KA)} f_T dt_2(KA).
\end{align*}
Therefore, the computed value of the frequency of Eq.~\eqref{eq:f_c} may be cast into the following form:
\begin{align*}
  f_c = \frac{1}{T_c}\int_{t_2^b(KA)}^{t_2^e(KA)} f_T dt_2(KA) 
  = \frac{1}{T_c}(t_2^e(KA) - t_2^b(KA))(f_0 + \delta f_0),
\end{align*}
where $f_0$ is the nominal value of the frequency of the transmitted signal, $\delta f_0$ is the frequency correction, which accounts for the offset of the frequency standard's output frequency from its nominal value and the instability of this signal over the duration of transmission.

The start and stop times of transmission of the signal by the spacecraft and its reception by the ground station are related by the light-time equation \cite{moyer2005formulation}:
\begin{align}
  \label{eq:1w_delay}
  \begin{split}
    t_3(ST) - t_2(KA) & = \frac{r_{23}}{c} + RLT_{23} + \\
    & + (ST - ET)_{t_3} - (KA - ET)_{t_2} + \\
    & + \frac{1}{c}(\Delta_A(t_3) + \Delta_{KA}(t_2)) + \\
    & + \Delta_t(\mathbf{r}_{23}) - \Delta_i(\mathbf{r}_{23}) + \\
    & + \tau_D,
  \end{split}
\end{align}
where:
\begin{itemize}
\item [$\mathbf{r}_{23}$] --- the difference between the spacecraft position vector at the time of signal transmission and the station position vector at the time of signal reception,
  $\mathbf{r}_{23} = \mathbf{r}_2 - \mathbf{r}_3$;
\item [$RLT_{23}$] --- the relativistic time delay of signal propagation due to the gravitational field, primarily that of the Sun;
\item [$(ST - ET)_{t_3}$] --- the difference between the station time scale and the ephemeris time scale computed at the time of reception, $t_3$;
\item [$(KA - ET)_{t_2}$] --- the difference between the station time scale and the ephemeris time scale computed at the time of transmission, $t_2$;
\item [$\Delta_A(t_3)$] --- the correction due to the receiving antenna phase center displacement relative to the station reference point, computed at the time of reception, $t_3$;
\item [$\Delta_{KA}(t_2)$] --- the correction due to the receiving antenna phase center displacement relative to the station reference point, computed at the time of transmission, $t_2$;
\item [$\Delta_t(\mathbf{r}_{23})$] --- the tropospheric signal delay computed for the spacecraft--station downlink;
\item [$\Delta_i(\mathbf{r}_{23})$] --- the signal delay due to charged particles,
computed for the spacecraft--station downlink;
\item [$\tau_D$] --- the signal delay due to the signal propagation in the station hardware.
\end{itemize}

The delay due to the ionosphere is assumed to be positive and, since we
are considering phase measurements, it is thus taken with the minus sign in Eq.~\eqref{eq:1w_delay}.

Using Eq.~\eqref{eq:1w_delay}, we obtain for the computed value of the frequency
observable:
\begin{align*}
  f_c = \frac{1}{T_c}&(f_0 + \delta f_0)\left(t_3^e(ST) - t_3^b(ST) - \right)\nonumber\\
  - &\left.\left[(t_3^e(ST) - t_2^e(KA)) - (t_3^b(ST) - t_2^b(KA))\right]\right),
\end{align*}
or, denoting by $\rho_{23}(t)$ the one-way delay of Eq.~\eqref{eq:1w_delay} related to the reception time $t$, we get:
\begin{align*}
  f_c = (f_0 + \delta f_0)\left(1 - \frac{\rho_{23}(t_3^e) - \rho_{23}(t_3^b)}{T_c}\right).
\end{align*}
Now, substituting $\rho_{23}$ here with its expression from Eq.~\eqref{eq:1w_delay}, we arrive at the following more detailed expression:
\begin{align}
  \label{eq:f_c_full}
  \begin{split}
    f_c = &(f_0 + \delta f_0)\left[1 - \frac{r_{23}(t_3^e) - r_{23}(t_3^b)}{c\cdot T_c}\right. -  \\
    - &\frac{1}{T_c}\left((ST - ET)_{t_3^e} - (ST - ET)_{t_3^b}\right) + \\
    + &\frac{1}{T_c}\left((KA - ET)_{t_2^e} - (KA - ET)_{t_2^b}\right) - \\
    - &\frac{1}{T_c}\left(RLT_{23}(t_3^e) - RLT_{23}(t_3^b)\right) - \\
    - &\frac{1}{T_c}\left(\frac{\Delta_A(t_3^e) - \Delta_A(t_3^b)}{c}\right) - \\
    - &\frac{1}{T_c}\left(\frac{\Delta_{KA}(t_2^e) - \Delta_{KA}(t_2^b)}{c} \right) - \\
    - &\frac{1}{T_c}\left(\Delta_t(t_3^e) - \Delta_t(t_3^b)\right) +\\
    + &\left.\frac{1}{T_c}\left(\Delta_i(t_3^e) - \Delta_i(t_3^b)\right)\right]. 
\end{split}
\end{align}

The terms in square brackets in Eq.~\eqref{eq:f_c_full} are the mean rates of change of the respective quantities over the considered time intervals. The most
interesting to us is the second term, which represents the doppler shift and is the origin of the largest uncertainty, and also the fourth one, which represents the influence of special and general relativity onto the spacecraft time scale. The latter is:
\begin{align*}
  \frac{1}{T_c}\left((KA - ET)_{t_3^e} - (KA - ET)_{t_3^b}\right) = 
  \frac{1}{T_c}\int_{t_2^b}^{t_2^e}\left(\frac{d\tau}{dt} - 1\right)dt,
\end{align*}
where $\tau = t(KA)$ is the proper time. The rate of change of the difference between the ephemeris time and the proper time is determined by the space-time metric and to order $1/c^2$ is given by:
\begin{align}
  \label{eq:dtau}
  \frac{d\tau}{dt} = 1 - (1+\varepsilon)\frac{U}{c^2} - \frac{1}{2}\frac{v^2}{c^2} + L,
\end{align}
where $U$ is the gravitational potential at the spacecraft position, $v$ is the spacecraft velocity, $L$ is a scale factor determined by the reference
frame used, and $\varepsilon$ is the general relativity violation parameter.

Using Eq.~\eqref{eq:dtau}, we obtain for the computed value of the received frequency:
\begin{align*}
  f_c = (f_0 + \delta f_0)\left[1 - \frac{r_{23}(t_3^e) - r_{23}(t_3^b)}{c\cdot T_c}
  - \frac{1+\varepsilon}{T_c}\int_{t_2^b}^{t_2^e}\frac{U}{c^2}dt - \right.\\
  \left.-\frac{1}{T_c}\int_{t_2^b}^{t_2^e}\frac{v^2}{2c^2}dt + \frac{T_t}{T_c}L -
  \frac{\Delta\rho_f}{T_c} \right],
\end{align*}
where $T_t = t_2^e - t_2^b$ is the transmission time interval in the ephemeris time scale and $\Delta\rho_f$ is the change of all the other corrections to the propagation delay of Eq.~\eqref{eq:1w_delay}, which can be computed accurately enough from the available data.
Now we cast the above equation into a simpler form:
\begin{align*}
  f_c = (f_0 + \delta f_0)\left[1 - \frac{r_{23}(t_3^e) - r_{23}(t_3^b)}{c\cdot T_c} - 
  \frac{\Delta\rho}{T_c}-  \right.\\
  \left.- \frac{T_t}{T_c}\left((1+\varepsilon)\frac{\overline{U}}{c^2} + 
  \frac{\overline{v^2}}{2c^2} - L\right)\right],
\end{align*}
where the average values of the gravitational potential and the spacecraft velocity are:
\begin{align*}
  T_t\frac{\overline{U}}{c^2} = \int_{t_2^b}^{t_2^e}\frac{U}{c^2}dt, \quad 
  T_t\frac{\overline{v^2}}{2c^2} = \int_{t_2^b}^{t_2^e}\frac{v^2}{2c^2}dt.
\end{align*}
Finally, eliminating the explicit dependance on the transmission time 
interval, we obtain the following equation for the computed value of the frequency of the signal received by the station:
\begin{align}
  \label{eq:f_c_main}
  f_c = (f_0 + \delta f_0)\left[1 - \frac{r_{23}(t_3^e) - r_{23}(t_3^b)}{c\cdot T_c} - 
  \frac{\Delta\rho}{T_c}\right]\times\nonumber\\
  \times\left[1 - \left((1+\varepsilon)\frac{\overline{U}}{c^2} + 
  \frac{\overline{v^2}}{2c^2} - L\right)\right].
\end{align}

Using Eq.~\eqref{eq:f_c_main}, one can obtain the computed value of the frequency observable given the spacecraft and tracking station trajectories.
The unknown values of $t_2$ and $\mathbf{r}_{KA}(t_2) = \mathbf{r}_2$
may be obtained iteratively, starting from the seed value of e.g. $t_2^{(0)} = t_3$ and using equations for $t_3(ST) - t_2$ similar to those of Eq.~\eqref{eq:1w_delay}
but not taking into account the spacecraft time-scale.

\section{Computed values of laser range observables}
The principle of laser ranging is based on measuring the total propagation delay of a signal transmitted at a known time:
\begin{align*}
  \rho_c = t_3(ST) - t_1(ST).
\end{align*}
This time difference conforms to the solution of the light-time equation, thus:
\begin{align}
  \label{eq:rho_c}
  \begin{split}
    \rho_c = &\frac{r_{12}}{c} + \frac{r_{23}}{c} + RLT_{12} + RLT_{23} + \\
    + &\frac{1}{c}\left(\Delta_{T}(\mathbf{r}_{12}) + \Delta_{R}(\mathbf{r}_{23})\right)
    + \frac{1}{c}\left(\Delta_{cm}(\mathbf{r}_{12}) + \Delta_{cm}(\mathbf{r}_{23})\right) + \\
    + &\Delta_t(\mathbf{r}_{12}) + \Delta_t(\mathbf{r}_{23}) 
    + \tau_U + \tau_D,
  \end{split}
\end{align}
where $\Delta_T$ and $\Delta_R$ are the corrections to the positions of the
transmitting and receiving antenna phase centers relative to the station's reference point,
$\Delta_{cm}(\mathbf{r}_{12})$ and $\Delta_{cm}(\mathbf{r}_{23})$ are
the corrections to the center-of-mass position for the direct and reflected beams, 
$\Delta_t$ is the correction for the tropospheric signal delay,
$\tau_T$ and $\tau_R$ are the corrections for delays in the transmitting and receiving hardware. 
The unknown time of reflection is computed iteratively starting from the seed value of $t_2^{(0)} = t_3$. 
Proceeding the same way as in the case of frequency measurements, we combine those terms which can be computed with the accuracy required by the experiment into a single term and denote it by $\Delta\rho_s$:
\begin{align}
  \label{eq:rho_main}
  \rho_c = &\frac{r_{12}}{c} + \frac{r_{23}}{c} + \Delta\rho_{s}.
\end{align}

\section{Joint analysis of the one-way frequency and two-way laser range
observables}
The reason for joint analysis of the two data types is obvious from the expressions
for their computed values  given in Eqs. \eqref{eq:f_c_main} and
\eqref{eq:rho_main}. Each of the two observables depends on the unknown values
of $r_{12}$ or $r_{23}$, which may be computed only by using orbital data and,
presumably, with  insufficient accuracy.

Suppose we have, as a result of simultaneous co-located observations, the
following two data sets: the set of one-way frequency measurements along
with their time tags:
\begin{align*}
  \mathbf{f}_o = \{f_o^1,\ldots,f_o^N\}, \quad \mathbf{t}_f = \{t_f^1,\ldots,t_f^N\},
\end{align*}
and the set of two-way delays obtained by laser ranging:
\begin{align*}
  \bm{\rho}_o = \{\rho_o^1,\ldots,\rho_o^M\}, \quad \mathbf{t}_{\rho} = \{t_{\rho}^1,\ldots,t_{\rho}^M\}.
\end{align*}
If the computed values of these observables were derived 
from the true orbit and with perfect modelling of every correction term,
they would differ from the observed values of these observables by random errors with zero mean:
\begin{align*}
  \mathbf{f}_o = \mathbf{f}_c(\mathbf{Q}^*) + \bm\varepsilon^*_f,\\
  \bm{\rho}_o = \bm{\rho}_c(\mathbf{Q}^*) + \bm\varepsilon^*_{\rho},
\end{align*}
where $\mathbf{Q}^*$ is the true vector of solve-for parameters, including those that determine the spacecraft trajectory and the correction terms of
Eqs.~\eqref{eq:1w_delay} and \eqref{eq:rho_c},
and $\bm\varepsilon^*_{f}$ and $\bm\varepsilon^*_{\rho}$ are random error vectors, $E(\bm\varepsilon^*_{f}) = E(\bm\varepsilon^*_{\rho}) =
0$. Suppose there exists another vector  $\mathbf{Q}$, which is different
from $\mathbf{Q}^*$ and fits the laser observations with a different set
of residuals:
\begin{align*}
  \bm{\rho}_o = \bm{\rho}_c(\mathbf{Q}) + \bm{\varepsilon}_{\rho},
  \quad E(\bm{\varepsilon}_{\rho})\approx 0.
\end{align*}

For a particular observed value of range the corresponding computed value, $\rho_c(\mathbf{Q})$, is known, in contrast to $\rho_c(\mathbf{Q^*})$. The computed value of the one-way delay, $r_{23}(t_{\rho}, \mathbf{Q})$, required to compute the one-way frequency observable, is also known. The accuracy of this computed frequency value depends, according to Eq.~\eqref{eq:f_c_main}, on the accuracy of the computed values of
$\Delta\rho(\mathbf{Q})$, $\overline{U}(\mathbf{Q})$,
$\overline{v^2}(\mathbf{Q})$, and
$(r_{23}(t_3^e, \mathbf{Q}) - r_{23}(t_3^b, \mathbf{Q}))/c$.
Let us consider the last term, 
$(r_{23}(t_3^e, \mathbf{Q}) - r_{23}(t_3^b, \mathbf{Q}))/c$,
in more detail since it is the origin of the largest part of error coming
from the uncertainties in the trajectory.

In order to estimate the errors involved, we take the difference of the previous two expressions and, using Eq.~\eqref{eq:rho_main} for $\bm{\rho}_c$, obtain:
\begin{align}
  &\left(\frac{r_{12}(\mathbf{Q}^*)}{c} + \frac{r_{23}(\mathbf{Q}^*)}{c}\right) - \left(\frac{r_{12}(\mathbf{Q})}{c} + \frac{r_{23}(\mathbf{Q})}{c}\right) = \nonumber\\
  &= \Delta\rho_s(\mathbf{Q}) - \Delta\rho_s(\mathbf{Q}^*) + 
    \varepsilon_{\rho} - \varepsilon_{\rho}^*,\nonumber\\
  \label{eq:err_diff}
  &\rho_{geom}(\mathbf{Q}^*) - \rho_{geom}(\mathbf{Q}) = \delta\rho_s + \delta\varepsilon_{\rho}.
\end{align}

The left-hand side of Eq.~\eqref{eq:err_diff} is the error in the computed two-way delay, while the right-hand side contains the sum of errors in the computed values of various corrections, $\delta\rho_s$, and the bias error, 
$\delta\varepsilon_\rho = \varepsilon_{\rho} - \varepsilon_{\rho}^*$, in
the fit of the range data with the model parameterized by $\mathbf{Q}$.

There is yet another contribution to the computed one-way delay error, i.e. that due to the uncertainty in the reflection time of the laser pulse off the spacecraft retroreflectors:
\begin{align}
  \label{eq:1w_err}
  \delta\left(\frac{r_{23}}{c}\right) = \frac{1}{2}\delta\rho_s +
  \delta\rho_{1w} + \frac{1}{2}\delta\varepsilon_{\rho}.
\end{align}
This error is due to the fact that for a given value of the geometric delay $(r_{12} + r_{23})/2$, the individual terms of the sum can be determined
only with some uncertainty.

Note that since Eq. \eqref{eq:f_c_main} involves the difference of two one-way delays,\textbar\ special attention must be paid to those terms that may change during the joint observation. For the errors in corrections, $\delta\rho_s$, these terms are the error in the tropospheric delay model, the error in the location of the spacecraft center-of-mass relative to its retroreflectors, the uncertainty in the ITRF coordinates of the ground station's reference point.

An estimate of the part of $\delta\rho_{1w}$ due to the uncertainty in the time of reflection of the signal off the spacecraft could be obtained by the following reasoning. Consider the equation $r_{12} + r_{23} = \rho\cdot c$, which represents
a 3-dimensional ellipsoid with the focal points at the positions of the station at the times of transmission and reception. In an Earth-centered
inertial  reference frame these positions can be considered error free, since the times of laser pulse transmissions are
accurately clocked and the reception times are computed using the transmission
times and even more accurately clocked round-trip light times. In the plane defined by points $\mathbf{r}_1$, $\mathbf{r}_2$ and $\mathbf{r}_3$ the spacecraft position at the reflection time can be specified by its anomaly, $\nu$. The error
in  $\nu$, which is due to inaccuracy of the reference orbit, results in the error in the computed one-way delay:
\begin{align*}
  \delta r_{23} = \frac{ae(1-e^2)\sin\nu}{(1+e\cos\nu)^2}\delta\nu,
\end{align*}
which can also be given in terms of the error in position on the ellipse, $\delta l$, as:
\begin{align*}
  \delta r_{23} = \frac{e\sin\nu}{1 + e\cos\nu}\delta l.
\end{align*}
We can simplify this equation using the fact that the eccentricity of the ellipse is small and thus obtain the upper limit
for the magnitude of this error:
\begin{align}
  \label{eq:err_ellip}
  \delta r_{23} = \frac{v_{ST}}{c}\delta l,
\end{align}
where we expressed the eccentricity in terms of the station velocity, $v_{ST}$.
When using Eq.~\eqref{eq:err_ellip} in \eqref{eq:f_c_main}, the difference between the spacecraft position errors can be replaced, for short accumulation times, by the velocity error $\delta v_{KA}$:
\begin{align*}
  \frac{\delta r_{23}(t_3^e) - \delta r_{23}(t_3^b)}{T_c} \leq 
  \frac{v_{ST}}{c}\delta v_{KA},
\end{align*}
Thus, we are now able to conclude that the contribution of the second term in the right-hand side of Eq.~\eqref{eq:1w_err} to the error in the computed value of the spacecraft velocity is of order $2\cdot 10^{-5}$~mm/s. 

Consider now  $\delta\varepsilon_{\rho}$,  i.e. the difference between the residuals of two fits, one with the solve-for parameters set to $\mathbf{Q}$ and the other with them set to $\mathbf{Q}^*$.
Since the same data is fitted in both cases, the random error in the data
cancels in the difference and does not contribute to $\delta\varepsilon_{\rho}$.
Hence, $\delta\varepsilon_{\rho}$ is the bias error of the computed values of $\rho_c(t, \mathbf{Q})$. 
If the expected value of errors is indeed zero, $\delta\varepsilon_{\rho}$ also characterizes the accuracy of approximation of the data  $\{\bm{\rho}_o, \mathbf{t}_{\rho}\}$ with the curve $\rho_c(t, \mathbf{Q})$. The error of $\delta\varepsilon_{\rho}$ turns
out to be the most difficult to estimate.

A seemingly obvious approach would be to choose a specific class of functions and try to find the function from this
class which best fits the available range measurements.
Then, in order to estimate the above error, one would have to assume that the true range is also represented by a function
from the chosen class. However, for example, if one tries to fit the data with polynomials of some order, higher order terms will contribute to the error, and this contribution would be impossible to estimate.

We suggest the following two approaches to estimating $\delta\varepsilon_{\rho}$. The first
is to try to generate an improved solution for the spacecraft trajectory for the time interval of the observation by accounting for unmodeled accelerations using the Gauss-Markov process \cite{tapley2004OD}. Apart from a data set of range measurements
and a characterization of its noise, the following
additional input data
are required to generate
such a solution:
\begin{itemize}
\item [$\mathbf{Q}_0$] --- the vector of initial values of parameters at time $t_0 = t_{\rho}^1$;
\item [$\mathbf{K}_{Q_0}$] --- a priori error covariance matrix $\mathbf{Q_0}$;
\item [$\bm{\sigma}_u$] --- characteristic values of unmodeled accelerations along the three axes (the standard deviation of the random part of the process);
\item [$\tau$] --- the characteristic correlation time of the random process
used to account for unmodeled accelerations.
\end{itemize}

If we succeed in generating such an improved solution, it gives us a new $\mathbf{Q}$
vector and a new error covariance matrix, $\mathbf{K}_{Q}$.
Then the error of $\delta\varepsilon_{\rho}$ at time $t_i$ can be estimated using the following
equation:
\begin{align*}
  \sigma^{2}_{\rho} = 
  \left(\frac{\partial\rho(t_i)}{\partial\mathbf{Q}}\right)
  \mathbf{K}_{Q}
  \left(\frac{\partial\rho(t_i)}{\partial\mathbf{Q}}\right)^{\mathsf{T}},
\end{align*}
Obviously, the reliability of this result depends on the validity of the assumptions
we made, i.e. that the initial guess values of parameters are accurate enough
and the characteristics of unmodeled accelerations are chosen properly. This is the main disadvantage of this approach.

An alternative approach, which allows us to avoid using the covariance analysis and introducing hard-to-verify assumptions, is to set an upper limit on
the one-way range error, $\delta\varepsilon_{\rho}/2$, equal to $\sigma^2_{\rho} = E(\varepsilon_{\rho}^2)/4$, i.e. to the error of a single range data point. This approach is fruitless for short data accumulation times but for time
intervals comparable to the duration of the observation it can give satisfactory results. An unfavorable feature of this approach is that it requires us to
deal with the problem of evaluating the integrals in Eq.~\eqref{eq: f_c_main}.

\section{Conclusions}

We have shown that in order to obtain the computed values of the one-way
frequency observables, one needs to accurately evaluate the difference of one-way delays. The latter may be computed from laser ranging data. The task
of obtaining one-way delays from two-way laser data is complicated due to
the existence of three kinds of error: the errors in the modeled corrections,
the error in the reflection time of the laser pulse from the spacecraft, and the error of fitting the spacecraft trajectory to the laser data. We showed
that the error due to fitting is the most difficult
 one to estimate and suggested two approaches to this problem.

\section*{Acknowledgements}
The work of MVZ is supported by the RSF grant 17-12-01488 (analysis of laser ranging and frequency measurements for RadioAstron gravitational redshift experiment). The work of DAL is supported by the RFBR grant 17-02-01298 (estimation of accuracy of gravitational redshift tests).
The RadioAstron project is led by the Astro Space Center of the Lebedev Physical Institute of the Russian Academy of Sciences and the Lavochkin Scientific and Production Association under a contract with the Russian Federal Space Agency, in collaboration with partner organizations in Russia and other countries.

\bibliographystyle{elsarticle-num-names}

\bibliography{grav}








\end{document}